# High Strain Rate Behaviour of Nano-quasicrystalline $Al_{93}Fe_3Cr_2Ti_2$ Alloy and Composites

*S. Pedrazzini[a], M. Galano[b], F. Audebert[b,c,d], P. Siegkas[e], R. Gerlach[f], V.L. Tagarielli[g], G. D. W. Smith[b]*

**Affiliations**
[a] Department of Materials, Imperial College, London, SW7 2AZ, London, UK.
[b] Department of Materials, University of Oxford, Parks Road OX1 3PH, Oxford, UK.
[c] Advanced Materials Group, INTECIN (CONCET-UBA), Facultad de Ingeniería, Universidad de Buenos Aires, Paseo Colón 850, Buenos Aires, 1063, Argentina.
[d] Department of Mechanical Engineering and Mathematical Sciences, Oxford Brookes University, Wheatley Campus, OX33 1HX, Oxford, UK.
[e] School of Science and Technology, Nottingham Trent University, 50 Shakespeare Street, Nottingham, NG14FQ, UK.
[f] Department of Engineering, University of Oxford, Parks Road, OX1 3PJ, Oxford, UK.
[g] Department of Aeronautics, Imperial College London, SW7 2AZ, London, UK.

**Abstract**
In the present work, we demonstrate for the first time the outstanding dynamic mechanical properties of nano-quasicrystalline $Al_{93}Fe_3Cr_2Ti_2$ at.% alloy and composites. Unlike most crystalline aluminium-based alloys, this alloy and composites exhibit substantial strain rate sensitivity and retain much of their ductility at high rates of strain. This opens new pathways for use in safety-critical materials where impact resistance is required.

**Main**

Quasicrystals were first observed in rapidly solidified Al-Mn-(Fe/Cr) alloys by Daniel Shechtman [1], who received the 2012 Nobel prize for chemistry for the paradigm-shifting discovery of this new state of matter [2]. Since their discovery, quasicrystals have been observed in a variety of systems including laboratory-made binary [3], ternary [4], quaternary alloys [5], soft matter [6] and naturally occurring minerals [7,8]. Quasicrystals have shown extremely versatile functional properties due to the ordered, non-translational nature of their quasi-periodic lattice [9], making them suitable for an incredibly wide range of applications such as superconductivity [10], photonics [11–13], sub-wavelength apertures for near-field microscopy and photolithography [14], coatings (due to a low-friction coefficient along aperiodic axes) [15], as well as outstanding mechanical strength under quasi-static loading conditions [16] and thermal stability up to 75% of the alloys homologous melting temperature [17]. Studies have analysed the active deformation mechanisms under quasi-static loading that result in excellent mechanical response at both at both ambient temperature [18] and high temperatures [19], while no work has been performed to date on the response of these materials at high strain rates. In mechanical studies, the quasicrystals themselves have been modelled primarily as hard, incoherent particles which strengthen the alloy by interfering with dislocation motion [18,19], although direct evidence of ductility within quasi-periodic lattices was also observed in some cases [20], while in other cases quasicrystals nucleation at dislocations was reported [21].

We studied the dynamic mechanical performance of a quaternary quasicrystalline $Al_{93}Fe_3Cr_2Ti_2$ at.% alloy (hereafter denoted QC) and the composites obtained by mixing 10 and 20 vol.% pure Al fibres into a matrix of QC alloy (referred to as QC-10 and QC-20 respectively). Rapid solidification is required for quasicrystals production in the selected alloy (atomisation has been previously shown to produce a nano-quasicrystalline structure [18], while spray forming produced a microstructure made of other nano-scale intermetallic phases [22]). Other methods can also be used to achieve rapid solidification, but atomisation is preferred due to the up-scaling potential to allow industrial production. The alloy atomisation conditions, consolidation methods and influence of processing parameters are available in a previous publication [18]. X-ray diffraction (XRD) was performed for the identification of bulk phases, using a Philips 1810 θ–2θ diffractometer with Cu-Kα radiation. Diffractograms were collected over 2θ angles from 20 to 100° (scattering vector Q between 1.4 – 4.0 Å$^{-1}$) with a tube voltage of 35 kV, current of 50 μA, scanning step size 0.02°. Samples were prepared by grinding extruded bars into powders to remove texture effects and ease phase identification. Diffractogram peaks were indexed using a combination of data sheets [23] and published papers [17]. Scanning Electron Microscopy (SEM) was performed on a JEOL 6500F microscope. A working distance of 15 mm was used, 300 pA and a voltage of 30 kV. Samples were prepared by cutting longitudinal sections of the extruded bars, grinding and polishing them using increasingly finer grades of diamond and finally 0.04 mm colloidal silica.

Transmission Electron Microscopy (TEM) was performed on a Philips CMS-X4 W-filament microscope. Samples were prepared with a Gatan dimple grinder and Gatan Precision Ion Polishing System. The quasicrystalline particles were too small for the smallest selected area aperture to be used for electron beam diffraction. Convergent beam diffraction (CBD) was performed instead using the smallest spot sizes, which when converged would allow diffraction patterns to be taken from microstructural features as small as ~40 nm. Manual measurements of particle sizes were performed on bright-field images for maximum accuracy over ~120 measurements, which were then plotted as a graph and fitted to a Gaussian distribution. Results are quoted as mean value ± full-width-half-maximum (FWHM) of the distribution. Energy dispersive X-ray spectroscopy (EDX) was used to measure chemical compositions of quasicrystals and their surrounding matrix with an Oxford Instruments EDX detector. Values quoted are mean ± standard deviation obtained over 10 measurements in randomly distributed locations on the 3mm TEM disc.

Tensile and compressive quasi-static tests were performed at a strain rate of $10^{-4}$ $s^{-1}$ at room temperature. Tensile quasi-static tests were performed by Westmoreland plc using an Instron servo-hydraulic mechanical testing machine and measuring strain by means of an extensometer. Compressive quasi-static tests were performed in-house using an Instron servo-hydraulic testing machine in displacement control. In compression cylindrical specimens were used, of diameter 2 mm and height 2 mm. The number of tests was limited by the volume of material available but at least 2-4 repeats of each test were performed. Quasi-static tests were highly repeatable, with scatter less than 1%. Dynamic tests were performed using a split-Hopkinson bar set-up available in the University of Oxford. Test pieces were marked using evenly spaced markers and the tests were filmed by means of a high-speed camera. Strain was measured through digital image correlation (using the marks as references). Between 2-4 repeats of each test were performed and representative graphs were selected and displayed. Dog-bone shaped threaded round-section samples of ASTM standard size E8/E 8M sub-size 4 (diameter 4 mm, gauge length 16 mm) were used in tension. Dynamic equilibrium in the tensile tests was achieved only in a small number of experiments, and only just prior to the specimen failure; for this reason the measured tensile stress versus strain curves are not presented, but only the measured tensile strength is reported; the average strain rates indicated are obtained by averaging the measured strain rate between the point of dynamic equilibrium and the time at which peak stress is achieved. Specimens of diameter 3 mm x 6 mm were used in compression. During dynamic compression tests, dynamic equilibrium was achieved towards the end of the elastic response.

Detailed microstructural characterisation was performed. Phase identification within samples of the QC alloy and QC-10 and QC-20 composites was performed by X-ray diffraction (XRD), displayed in Figure 1. The α-Al peaks corresponding to the (111), (200), (220) and (311) were indexed [23] alongside the icosahedral phase, with reflections visible at angles [17] 2θ ~ 23°, 41°, 43°, 62°, 74° (scattering vectors Q=1.6, 2.7, 2.8, 3.6 and 3.9 Å$^{-1}$), indicating that the overall microstructure was constituted solely of those two phases (or that if other phases were present, their volume fraction did not exceed the XRD detection limit of ~4%). The phases detected are consistent with previous XRD studies on gas atomised quasicrystalline Al-Fe-Cr-Ti alloy, who all reported a microstructure composed mainly of an FCC α-Al matrix and icosahedral phase, although occasionally, if the solidification rate was varied, other intermetallic phases were detected. Todd et al. reported the presence of $Al_{23}Ti_9$ [24] which is a metastable phase also seen in the work of Inoue and Kimura [25]. Garcia Escorial et al. [26] and Yamasaki et al. [27] also showed Al-Ti intermetallics in their diffractograms, however this time it was stable $Al_3Ti$ intermetallics. Metastable distorted $Al_{13}(Fe-Cr)_{2-4}$ also known as θ-phase was found in $Al_{93}Fe_3Cr_2Ti_2$ alloy produced at a higher cooling rate by melt spinning by Galano et al. [28] and $Al_3Ti$, $Al_{13}Fe_4$ and $Al_{13}Cr_2$ intermetallics. Garcia Escorial et al. showed that the same alloy produced by spray forming (lower cooling rate) contained stable $Al_3Ti$, $Al_{13}Fe_4$ and $Al_{13}Cr_2$ intermetallic phases [26]. All the aforementioned intermetallic phases were taken into consideration when processing the diffractograms in Figure 1, though only α-Al and the icosahedral quasicrystalline phase were confirmed. Peak intensity comparisons were performed on the α-Al peaks, comparing them to an untextured powder sample, indicating that the thermo-mechanical production of all samples through extrusion induced a ~60% texture along the (220) direction (preferred grain orientation), which is itself an unusual finding for an aluminium alloy. This was measured by peak-intensity comparison, using the textures bar and untextured powder produced by grinding the bar.

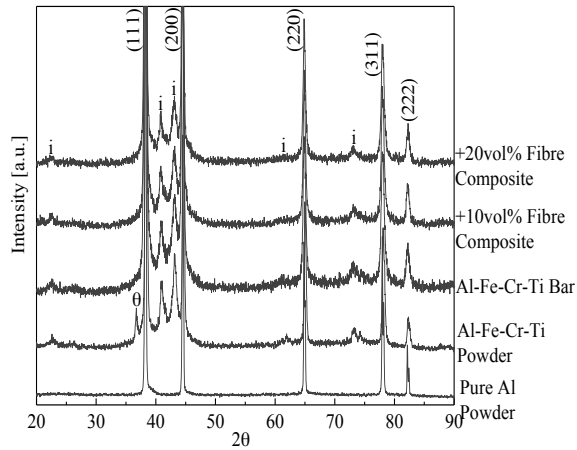

***Figure 1:*** *X-ray diffractogram of the QC, QC-10 and QC20 samples confirming the presence of icosahedral phase, α-Al and the same phases were present in the atomised powders before extrusion, with the exception of θ which decomposed during extrusion.*

Figure 2 (a-c) shows SEM micrographs of longitudinal sections of the extruded QC, QC-10 and QC-20 bars, revealing a microstructure made of deformed well-bonded powder particles, with no evidence of residual porosity remaining after extrusion. Some randomly distributed pure Al contamination was detected in the QC sample, which likely originates from the standard industrial practice of cleaning the atomiser with pure Al in between production of different alloys. This contamination was quantified at ~0.35 vol.% by synchrotron X-ray tomography [18], which confirmed SEM-based observations. The extruded microstructure of the QC-10 and QC-20 composite bars consists of a quasicrystalline alloy matrix with 10-20 vol% pure Al fibres. Fibres are made from extruded <25 μm diameter pure Al powder particles, which during extrusion were deformed into fibres <200 μm in length. The sample longitudinal section and cross section show a random distribution of fibres within the composites, with no observed powder clustering. Grain size and texture was previously measured by EBSD [18] giving a mean planar grain diameter of $d_s$= 0.97±0.3 μm (mean ± standard deviation).

TEM micrographs of the icosahedral quasicrystalline particles and their respective 5-fold symmetry convergent beam diffraction pattern (CBDP) are shown in Figure 2d. A bimodal distribution of well-dispersed icosahedral quasicrystalline phases was identified with mean spherical radii $r_1$=43 ± 6 nm and $r_2$=129 ± 6 nm. The volume fraction of quasicrystals in the FCC-Al matrix was 0.41, slightly lower than the 0.45 value measured by Inoue and Kimura, who produced the same alloy by melt spinning instead of atomisation [25]. Quasicrystal composition measured by EDX was $Al_{87.8}Fe_{4.6}Cr_{4.2}Ti_{3.4}$ at.%. The composition of the α-Al matrix between quasicrystalline particles was $Al_{98.1}Fe_{0.3}Cr_{0.3}Ti_{1.3}$ at.%. All matrix values are higher than the equilibrium solid solubility of each element in an α-Al matrix, particularly titanium [29], a likely result of the rapid solidification production route.

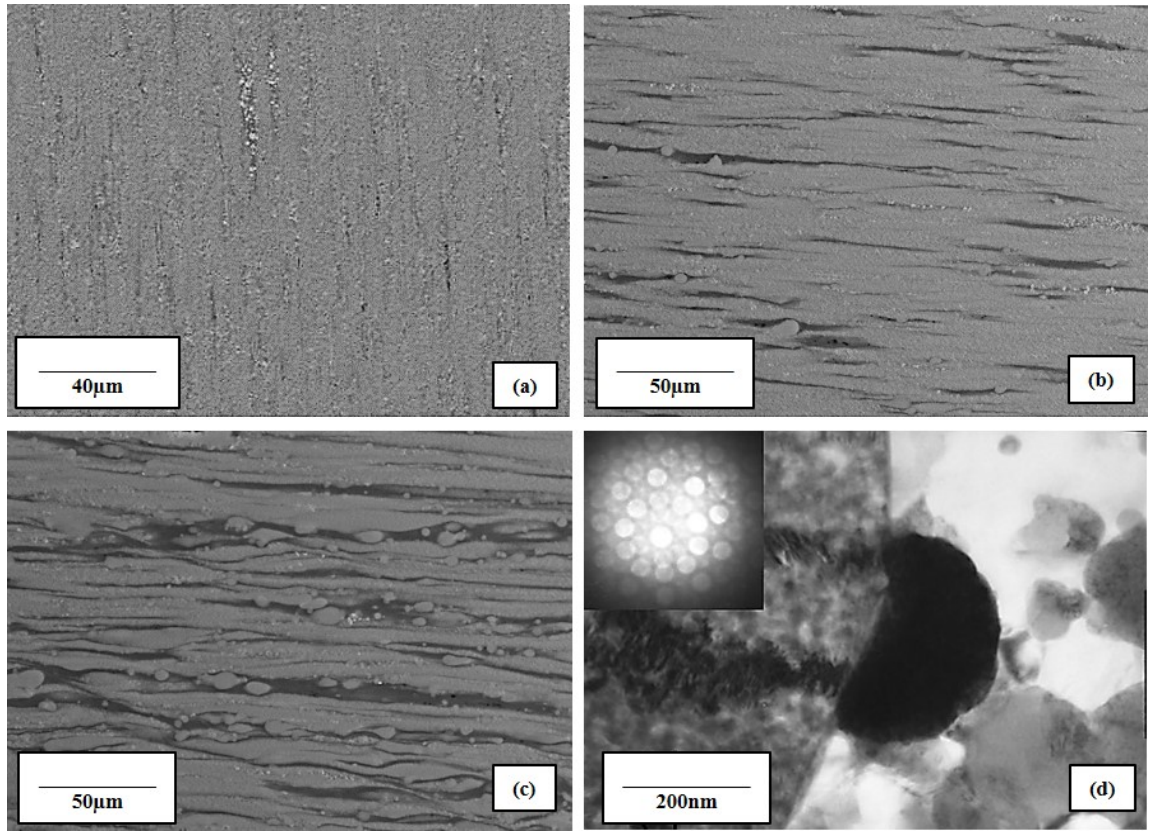

*Figure 2: (a) backscattered low magnification overview of the QC bar, (b) the QC-10 bar, (c) the QC-20 bar, (d) bright field TEM micrograph and CBDP showing the characteristic 5-fold symmetry of the icosahedral phase.*

Figure 3 shows compressive true stress-strain curves of the as-extruded QC alloy and QC-10 and QC-20 composites at strain rates ranging from $10^{-4}$ s$^{-1}$ to 2840 s$^{-1}$, with a summary of all tests performed shown in Table 1. In compression, the QC alloy strength was 0.74±0.10 GPa at slow strain rate ($10^{-4}$ s$^{-1}$) and 0.90±0.25 GPa at fast strain rates (~3x10$^3$ s$^{-1}$). Strength of the QC alloy is strain-rate dependent, increasing by ~20% from quasi-static to dynamic loading conditions, though the dynamic ductility, of order 5-6%, was reduced compared to the quasi-static case (10-20%). Localised plastic deformation initiated along shear bands inclined at approximately 45° from the direction of loading, corresponding to a plateau in the true stress-strain curves, shown in Figure 3. The shear bands width progressively increased to 2-3 mm and quickly evolved into a macroscopic crack, causing ultimate failure of the specimen (see insert of Figure 3). This behaviour was observed both during quasi-static and dynamic tests, though the width of shear bands was narrower after dynamic tests.

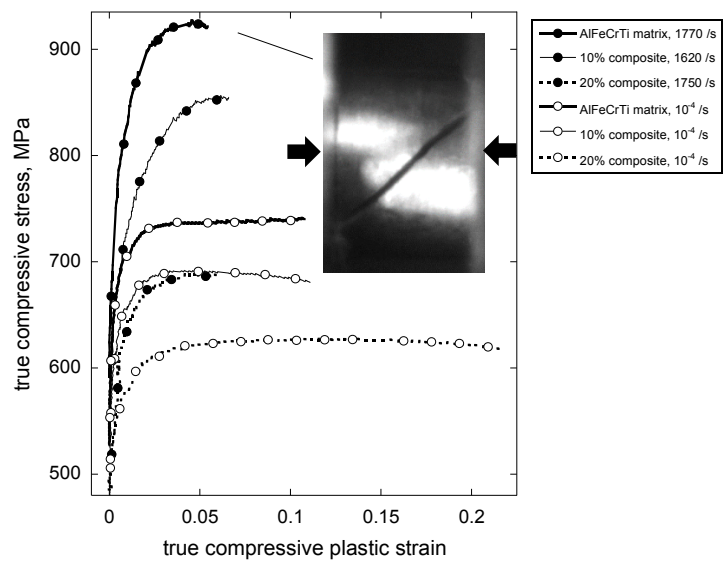

***Figure 3:*** *compressive true stress-strain curves of the QC alloy, QC-10 and QC-20 composites under quasi-static (1x10$^{-3}$ s$^{-1}$) and dynamic (3x10$^3$ s$^{-1}$) loading conditions.*

In tension, the strength of the QC alloy was 0.66±0.1 GPa at slow strain rate ($10^{-4}$ s$^{-1}$) and 0.97±0.2 GPa at fast strain rates ($3 \times 10^3$ s$^{-1}$). Strain-rate dependency was more marked in tension, with a strength increase of 32% from quasi-static to dynamic loading conditions, though the dynamic ductility, of order 5-6%, was substantially reduced compared to the quasi-static case (10-20%). A summary of all collected mechanical data is shown in Table 1. In quasi-static tension, the QC alloy displayed an initial hardening phase followed by brittle fracture along a plane perpendicular to the loading direction, at plastic strains up to ~6%. In dynamic tension, the materials response was brittle and specimens failed by a transverse crack at the end of the elastic response, showing only limited plastic deformation and no visible necking. Repeated dynamic tests revealed wide scatter in the measured strength, of the order ±0.05 GPa. Failure was occasionally initiated by large pores, consequences of imperfect manufacturing. The moduli of the three materials were approximately insensitive to the direction of loading (tensile or compressive), though the compressive strength marginally outperformed the tensile strength independently of the strain rate. As expected from the conventional rule of mixtures, strength decreased linearly with increasing volume fraction of fibres, though the ductility and toughness required at least 20 vol% fibres in order to improve.

| Test # | Type (T/C) | Material | Strain rate, /s | Strength, MPa |
|---|---|---|---|---|
| 1 | C | QC | 0.0001 | 740 |
| 2 | C | QC-10 | 0.0001 | 690 |
| 3 | C | QC-20 | 0.0001 | 626 |
| 4 | T | QC | 0.0001 | 660 |
| 5 | T | QC-10 | 0.0001 | 605 |
| 6 | T | QC-20 | 0.0001 | 530 |
| 7 | C | QC | 1770 | 926 |
| 8 | C | QC | 2840 | 878 |
| 9 | T | QC | 1490 | 943 |
| 10 | T | QC | 2720 | 998 |
| 11 | C | QC-10 | 1620 | 856 |
| 12 | T | QC-10 | 1530 | 822 |
| 13 | C | QC-20 | 1750 | 689 |
| 14 | T | QC-20 | 1740 | 754 |

*Table 1: Summary of all tests performed in this study and relevant measurements.*

High strain rate tests were performed and the dynamic flow stress in tension measured for the pure alloy bar was ~1.00 GPa, while the composites showed 0.82 GPa (10 vol.% Al) and 0.75 GPa (20 vol.% Al). When divided by the density, their specific strength was measured to be 0.35 GPa/gcm$^{-3}$, 0.28 GPa/gcm$^{-3}$ and 0.26 GPa/gcm$^{-3}$ respectively. Values for commercially available Al alloys tested at similar strain rates were obtained from the literature and used as comparison, as shown in Figure 4. Tests were found on 7000 series Al alloys performed by Chen *et al.* [30] at 1000s$^{-1}$ strain rate. They found specific dynamic flow stresses of 0.10-0.18 GPa/gcm$^{-3}$. Reyes et al performed tests at 1300s$^{-1}$ strain rate and found that 7000-series alloys had a specific dynamic flow stress of 0.14-0.15 GPa/gcm$^{-3}$ [31]. Ti6Al4V tested at 3500s$^{-1}$ and had a specific flow stress of 0.42 GPa/gcm$^{-3}$ [32]. Cited data is summarised in Figure 4. While commercially available Al-based alloys display negligible strain rate sensitivity, the QC alloy under investigation becomes substantially stronger when tested at high strain rates. The dynamic strength of the alloy and its composites is 100-200% higher than that of commercial Al-alloys, and not far from that of the Ti6Al4V alloy, shown for comparison. The outstanding impact strength of the QC alloy and composites open new applications for safety-critical materials for impact resistant engineering components.

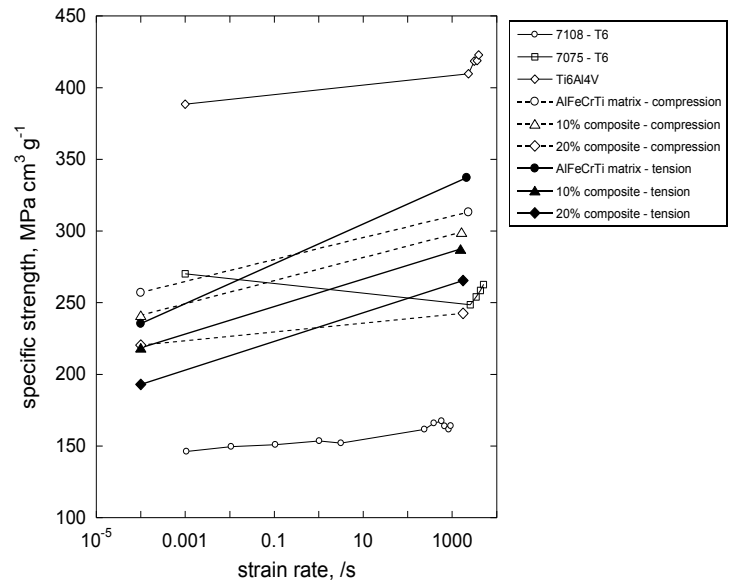

*Figure 4:* *Specific strength as a function of the imposed strain rate; the tensile and compressive responses of the* $Al_{93}Fe_3Cr_2Ti_2$ *alloy and its composites in comparison with the compressive strength of commercial high-strength AA7075-T6 [32], AA7108-T6 [31] and Ti6Al4V [32] alloys for dynamic applications.*

**Acknowledgements**

The authors would like to thank Mr C. Downing for help with sample preparation and operation of a JEOL 6500 SEM and ALPOCO Ltd. who provided the powders. EPSRC (Project EP/E040608/1) provided financial support.

**Data Availability**

The data that support the findings of this study will be made fully available upon reasonable request to the authors. A pre-print of this publication is available on ArXiv.